\documentclass[%
 aip,
 amsmath,amssymb,
 reprint,%
]{revtex4-1}

\usepackage{graphicx}
\usepackage{dcolumn}
\usepackage{bm}

\usepackage[utf8]{inputenc}
\usepackage[T1]{fontenc}
\usepackage{mathptmx}
\usepackage{etoolbox}

\makeatletter
\def\@email#1#2{%
 \endgroup
 \patchcmd{\titleblock@produce}
  {\frontmatter@RRAPformat}
  {\frontmatter@RRAPformat{\produce@RRAP{*#1\href{mailto:#2}{#2}}}\frontmatter@RRAPformat}
  {}{}
}%
\makeatother
\begin{document}


\title[Spontaneous and stimulated Raman microscopy signals]{Theoretical estimation of stimulated and spontaneous Raman signals in Raman microscopy}
\author{Yasuyuki Ozeki}
\affiliation{Research Center for Advanced Science and Technology, The University of Tokyo\\ 4-6-1 Komaba, Meguro-ku, Tokyo 153-8904, Japan.
}%

\date{\today}

\begin{abstract}
  As a highly sensitive vibrational imaging method, stimulated Raman scattering (SRS) microscopy is finding various applications, while its theoretical treatment seems still under development. Here we present a theoretical estimation of spontaneous Raman signal and SRS signal from Raman scattering cross section, irrespective of the numerical aperture of the objective lens. We confirm a reasonable agreement between our theory with a recently proposed treatment based on the stimulated Raman scattering cross section. Furthermore, we point out that the acceleration factor of SRS can be interpreted as the number of Stokes photons in the vibrational coherence time.
\end{abstract}

\maketitle

\section{Introduction}
Since its first demonstrations\cite{freudiger2008label, nandakumar2009vibrational, ozeki2009analysis}, stimulated Raman scattering (SRS) microscopy has evolved as a powerful imaging method providing molecular-vibrational contrasts\cite{cheng2021stimulated}. It has found various applications in biomedicine and materials science\cite{cheng2021stimulated}.
Nevertheless, a complete theoretical treatment of SRS seems still under development. The physics of SRS is understood as a third order nonlinear process\cite{yariv1989quantum, rigneault2018tutorial}. SRS can be also understood as stimulated emission process, which gives rise to an acceleration of Raman process by the number of Stokes photons\cite{boyd2020nonlinear, rigneault2022sensitivity}. Wei and Gao proposed stimulated Raman cross section to quantify the SRS process in the view of an interaction between two-color light and single molecules\cite{gao2023absolute}, and derived a relation between spontanous Raman cross section and SRS cross section\cite{min2023raman}. In this way, various pictures and interpretations of SRS process exist. However, the consistency among these pictures is yet to be thoroughly explored.

In this paper, we present a theoretical model to estimate spontaneous Raman and SRS signals from the cross section of spontaneous Raman scattering, irrespective of the focusing numerical aperture (NA) by assuming a homogeneous sample and the acceleration factor of SRS proposed previously by us\cite{rigneault2022sensitivity}. We demonstrate a reasonable agreement between our theory with that based on SRS cross section\cite{min2023raman}. We also point out that the acceleration factor of SRS can be interpreted as the number of Stokes photons in the vibrational coherence time. This work provides not only a useful equation for experimentalists but also an alternative view of the physics of SRS process.

\section{Theory}
We first derive the number of photons of spontaneous Raman signal and that of SRS signal under microscopic detection of a spatially homogenous sample. A part of derivation is described previously \cite{rigneault2022sensitivity}, but slightly adjusted. Then we connect SRS cross section\cite{gao2023absolute} to spontaneous Raman cross section to test our formulation in comparison with a rigorous theory based on the Raman interaction with single molecules\cite{min2023raman}.

\subsection{Spontaneous Raman signal from a homogeneous sample}
Here we derive the number of photons $m_\mathrm R$ of spontaneous Raman scattering in Raman microscopy. We assume that spontaneous Raman scattering of a spatially homogeneous sample is measured with a focused beam, whose entire focus is inside the sample. We also assume that the pump beam is focused with an objective lens using its full NA, and that the Raman scattering within the NA will be detected through a confocal pinhole, which limits the measurement region to a certain depth of focus. As a result, we obtain
\begin{equation}
    m_\mathrm R = \left(\frac{\mathrm d\sigma_\mathrm R}{\mathrm d\Omega}\right)_{\parallel + \perp} \Omega_\mathrm{OB} C \Delta z_\mathrm p \frac{U_\mathrm p}{\hbar \omega_\mathrm p}\frac{\hbar \omega_\mathrm p}{\hbar \omega_\mathrm S},
\end{equation}
where $(\mathrm d\sigma_R/\mathrm d\Omega)_{\parallel + \perp} [\mathrm m^2/\mathrm{Sr}]$ is the differential cross section (originally denoted as $\sigma_\mathrm R$ in our previous treatment\cite{rigneault2022sensitivity}) of both polarized Raman scattering and depolarized Raman scattering, $\Omega_\mathrm{OB}$ is the solid angle of the objective lens used for detecting the Raman scattering, $C$ is the concentration of the molecule, $\Delta z_\mathrm p$ is the depth of focus of the pump beam, $U_\mathrm p$ is the energy of pump beam, which can be thought as ether the pump beam power times the duration of continuous-wave irradiation or the pump energy in pulsed irradiation, $\hbar$ is the Planck constant divided by $2\pi$, $\omega_\mathrm p$ is the optical frequency of pump pulses, and $\omega_\mathrm S$ is the optical frequency of Stokes pulses. The last term (i.e., $\hbar\omega_\mathrm p / \hbar\omega_\mathrm S$), which was neglected in our previous treatment\cite{rigneault2022sensitivity}, accounts for the fact that the Raman cross section is defined not as the number but the energy of scattered photons\cite{min2023raman}. We then apply the following approximation:
\begin{align}
\Omega_\mathrm{OB} &= 2\pi(1-\cos \theta)= 4\pi\sin^2(\theta/2)
\sim \frac{\pi \mathrm{NA}^2}{n^2},
\end{align}
where $\theta$ is the acceptance angle of the objective lens and $n$ is the refractive index. We also estimate the depth of focus (see Appendix A for derivation) as
\begin{equation}
  \Delta z_\mathrm p = \frac{2n\lambda_\mathrm p}{\mathrm{NA}^2},  
\end{equation}
    where $\lambda_\mathrm p$ is the pump wavelength. By using eqs. (2) and (3), we obtain
\begin{equation}
\begin{aligned}
    m_\mathrm R 
    &= \left(\frac{\mathrm d\sigma_\mathrm R}{\mathrm d\Omega}\right)_{\parallel + \perp} \frac {2\pi \lambda_\mathrm p}{n}C \frac{U_\mathrm p}{\hbar \omega_\mathrm S}
\end{aligned}
\end{equation}
This equation connects the differential Raman cross section to the number of photons in Raman microscopy. It also shows that the signal is independent of NA\cite{rigneault2022sensitivity}. This is because $\Omega_\mathrm{OB}$ is proportional to $\mathrm {NA}^2$, while $\Delta z$ is proportional to $\mathrm{NA}^{-2}$ as shown in eq. (2) and eq. (3).

Furthermore, we use the following relationship to convert the differential Raman cross section to the total Raman cross section \cite{myers1987resonance, min2023raman}
\begin{equation}
    \sigma_\mathrm R = \frac {8\pi}{3}\frac{1+2\rho}{1+\rho}\left(\frac{\mathrm d\sigma_\mathrm R}{\mathrm d\Omega}\right)_{\parallel+\perp},
\end{equation}
where $\rho$ denotes the Raman depolarization ratio. In eq. (5), the polarized Raman cross-section is derived from the sum of polarized and depolarized Raman cross-sections by multiplying $(1+\rho)^{-1}$. Then, the total cross-section is obtained by multiplying with $1 + 2\rho$, considering that depolarized Raman scattering can occur from dipoles oscillating in two directions perpendicular to the pump polarization. The factor $8\pi/3 = 4\pi \times 2/3$ corresponds to the solid angle of dipole radiation of Raman scattering\cite{porto1966angular}, which does not occur in the direction parallel to the dipole oscillation \cite{feynman1965feynman}. From eqs. (4) and (5), we obtain 
\begin{equation}
    m_\mathrm R = \frac{3}{4} \frac{1+\rho}{1+2\rho}\sigma_\mathrm R \frac {\lambda_\mathrm p} n C \frac {U_\mathrm p}{\hbar \omega_\mathrm S}.
\end{equation}
This equation quantitatively connects the Raman microscopy signal of a homogeneous sample with the Raman cross section. Interestingly, $m_\mathrm R$ is equivalent to the Raman scattering collected in all the solid angle from a homogeneous sample with a thickness of $(3(1+\rho)/4(1+2\rho))(\lambda_\mathrm p/n)$, which is on the order of a wavelength. 

\subsection{Stimulated Raman signal from a homogeneous sample}
Here we derive the number of photons of SRS signal according to our previous approach\cite{rigneault2022sensitivity}, where we assume that the probability of SRS is accelerated by the number of Stokes photons, and that stimulated emission can occur only when the frequency spectrum of the Stokes photons matches the spontaneous Raman spectrum. 
This leads to the reduced acceleration efficiency of SRS by a factor of $1/\gamma \Delta t$, where $\gamma$ is the damping constant of the molecular vibration, and $\Delta t$ is the time duration of optical pulses with a rectangular waveform. Another important factor that was not considered previously is that SRS can accelerate the Raman process in the same polarization as Stokes photons. Assuming that pump and Stokes beams have the same polarization, SRS can only accelerate polarized Raman scattering, whose number of photons is $m_\mathrm R/(1+\rho)$. Hence, the number of SRS photons is given by
\begin{equation}
    m_\mathrm{SRS} = \frac{m_\mathrm R}{1+\rho}\frac{U_\mathrm S}{\gamma \Delta t \hbar\omega_\mathrm S}
    =\frac{3\sigma_\mathrm R \lambda_\mathrm p C U_\mathrm p U_\mathrm S}{4(1+2\rho)n\gamma \Delta t(\hbar\omega_\mathrm S)^2 }.
\end{equation}

\subsection{SRS cross section}
According to the definition of SRS cross section\cite{gao2023absolute}, $R_\mathrm{SRS}$ can be described as
\begin{equation}
        R_\mathrm {SRS} =\frac{m_\mathrm{SRS}}{\Delta t}
=\sigma_\mathrm{SRS}\phi_\mathrm p\phi_\mathrm S S_\mathrm p\Delta z_\mathrm p,
\end{equation}
where $\rho_\mathrm{SRS}$ is the SRS cross section, $\phi_\mathrm p$ and $\phi_\mathrm S$ are the photon flux of pump and Stokes pulses, respectively, and $S_\mathrm p$ is the focus area of pump pulses. The photon fluxes are given as
\begin{align}
\phi_\mathrm p &= \frac{U_\mathrm p}{\hbar \omega _\mathrm p \Delta t S_\mathrm p},    \\
    \phi_\mathrm S &= \frac{U_\mathrm S}{\hbar \omega_\mathrm S \Delta t S_\mathrm S},
\end{align}
where $S_\mathrm S$ are the focus area of Stokes pulses, respectively.
From eqs. (7)-(10), we obtain
\begin{equation}
\begin{aligned}
    R_\mathrm{SRS} 
    &= \frac{3\pi\sigma_\mathrm R cC\phi_\mathrm p\phi_\mathrm S S_\mathrm p S_\mathrm S}{2(1+2\rho)\gamma\omega_\mathrm S},
\end{aligned}
\end{equation}
where $c$ is the speed of light, and we used $2\pi nc = \omega_\mathrm p \lambda_\mathrm p$. By comparing eq. (8) with eq. (11), we get
\begin{equation}
    \sigma_\mathrm{SRS} = \frac{3\pi\sigma_\mathrm R cS_\mathrm S}{2(1+2\rho)\gamma\omega_\mathrm S\Delta z_\mathrm p}.
\end{equation}

Here, we estimate the focal volume of the Stokes beam (see Appendix B for derivation) as
\begin{equation}
S_\mathrm S=(1/\pi)(\lambda_\mathrm S/\mathrm{NA})^2, 
\end{equation}
where $\lambda_\mathrm S$ is the wavelength of the Stokes pulses. Also we use the relationship between the vibrational damping constant $\gamma$ and the vibrational linewidth $\Gamma$, $\gamma = \Gamma/2$. From eqs. (3), (12), and (13), we get
\begin{equation}
    \sigma_\mathrm{SRS}
    =\frac{3\sigma_\mathrm Rc\lambda_\mathrm S^2}{2(1+2\rho)\Gamma \omega_\mathrm S n \lambda_p}.
\end{equation}
By using $\lambda_\mathrm S = 2\pi c/n\omega_\mathrm S$, we obtain
\begin{equation}
\sigma_\mathrm{SRS}
    =\frac{3\pi\sigma_\mathrm Rc^2\omega_\mathrm p}{(1+2\rho)\Gamma \omega_\mathrm S^3}.
\end{equation}
With an implicit assumption of $\rho=1$ by Min and Gao\cite{min2023raman}, this leads to  
\begin{equation}
\sigma_\mathrm R = \frac {\Gamma \omega_\mathrm S^3}{\pi c^2\omega_\mathrm p}\sigma_\mathrm{SRS}.
\end{equation}
This equation gives two times larger spontaneous Raman scattering cross section compared with eq. (16c) in Ref. \cite{min2023raman}. The origin of this discrepancy is not clear at the moment, and will be discussed in Section III. Nevertheless, eq. (16) shows a reasonable consistency between two theories, one based on the response of single molecule\cite{gao2023absolute} and the other based on the response of a homogeneous medium\cite{rigneault2022sensitivity}. Importantly, we obtain the same dependence on $\Gamma$, $\omega_\mathrm S$, and $\omega_\mathrm p$. In the derivation, we can see that the factor of $\omega_\mathrm S^3/\omega_\mathrm p$ comes from eqs. (1) and (13), i.e., the spontaneous Raman scattering cross section is defined as the energy of scattered photon, and the focal area of Stokes photon is proportional to $\lambda_\mathrm S^2$.
\section{Discussion}
\subsection{Acceleration factor in the SRS process can be interpreted in several ways}
As described above, we previously obtained the acceleration factor of SRS process by considering that only a portion of a vibrational spectrum can be detected by SRS. Here, we point out here that the acceleration factor can be interpreted as the number of photons in the vibrational coherence time, which  is given by $\tau_\mathrm v= 1/\gamma$. The acceleration factor is $m_\mathrm S /\gamma\Delta t = m_\mathrm S \tau_\mathrm v / \Delta t$, where $m_\mathrm S$ is the number of Stokes photons. This interpretation is useful when SRS is excited with long pulses or continuous-wave (CW) beams, where the number of photons in the vibrational coherence time is smaller than the number of photons in the entire beam. 
These pictures are consistent with the model based on SRS cross section\cite{min2023raman}, where the vibrational spectral peak is inversely proportional to  the vibrational linewidth. In this way, these interpretations are consistent with each other.

\subsection{Dependence on NA}
Equation (7) shows that the acceleration factor does not depend on the focusing condition. This looks contradictory to eq. (8), which indicates that SRS rate depends on the photon flux of Stokes pulses. Instead, this is the consequence of the microscopic measurement of a homogeneous medium, where the focal area is proportional to $\mathrm{NA}^{-2}$ and the focal depth is proportional to $\mathrm{NA}^{-2}$. Thus the photon flux of pump and Stokes pulses is proportional to $\mathrm{NA}^{2}$, which results in the SRS rate per single molecule of $\propto \mathrm {NA}^4$. This cancels out the number of molecules in the focal volume of $\propto \mathrm{NA}^{-4}$, leading to the NA-independent SRS signal intensity. In turn, when the size of samples is smaller than focal volume, SRS signal is proportional to $\mathrm {NA}^4$, and thus the use of a high-NA objective lens is crucial for realizing sensitive measurement.

The independence of NA originates from the assumption that the focusing NA and the detection NA are the same in typical Raman microscopy. Even in SRS microscopy, where the detection NA is typically larger than the focusing NA to minimize unwanted cross-phase-modulation artifact, the modulation transfer from either pump or Stokes beam to the other beam is detected. Therefore, the effective detection NA is the same as the focusing NA. With this assumption, we can consider the Raman interaction in the single spatial mode, i.e., the focused beam. This greatly simplifies the analysis to derive the number of photons of spontanous Raman and SRS in eqs. (6) and (7). This is in contrast to typical treatment considering the Raman scattering from a collimated beam into multiple spatial modes that corresponds to the dipole radiation. This can explain the reason why typical spontaneous Raman spectroscopy uses an excitation beam loosely focused in the sample and Raman scattering is detected with a larger NA to increase the number of photons to be detected.
The independence of NA originates from the assumption that the focusing NA and the detection NA are the same in typical Raman microscopy. Even in SRS microscopy, where the detection NA is typically larger than the focusing NA to minimize unwanted cross-phase-modulation artifact, the modulation transfer from either pump or Stokes beam to the other beam is detected. Therefore, the effective detection NA is the same as the focusing NA. With this assumption, we can consider the Raman interaction in the single spatial mode, i.e., the focused beam. This greatly simplifies the analysis to derive the number of photons of spontanous Raman and SRS in eqs. (6) and (7). This is in contrast to typical treatment considering the Raman scattering from a collimated beam into multiple spatial modes that corresponds to the dipole radiation. This can explain the reason why in typical spontaneous Raman spectroscopy an excitation beam is loosely focused to the sample and the Raman scattering is detected with a larger NA to increase the number of photons to be detected.

\subsection{Uncertainties of the present analysis}
Physics can be understood by several pictures, and different interpretations lead to deeper understanding. For this, it is crucial that different interpretations lead to the same conclusion. In Section II, to test our thoretical estimation assuming a homogeneous sample, we derived the relationship between the spontaneous and stimulated Raman cross sections, and compared the result with the rigorous analysis based on the response of single molecules\cite{min2023raman}. Even with a modified estimation of the focal area in eq. (13), there still exist two times difference between them. This can be considered as a reasonable agreement, while the origin of this discrepancy is not clear. Possible origins for the discrepancy include the estimation of the SRS acceleration efficiency $1/\gamma\Delta t$, the focal area $S_\mathrm S$, and the assumption of single molecule or homogeneous medium. Note that with our original estimation of the focal area \cite{rigneault2022sensitivity}, $S_\mathrm{prev}=(2\pi)^{-1}(\lambda/\mathrm{NA})^2$,  the discrepancy becomes four times. Another possible origin is that the focused Stokes beam in SRS has an intensity distribution while vacuum fluctuation is spatially homogeneous.

\section{Conclusion}
We have presented the theoretical estimation of spontaneous Raman scattering and SRS signals in the microscope setting, showing that spontaneous Raman and SRS signals of homogeneous samples do not depend on the focusing NA. We confirmed a reasonable agreement between our theory and that based on SRS cross section. We also described our interpretation of the acceleration factor of SRS, i.e., SRS is accelerated by the number of photons of Stokes pulses in the vibrational coherence time. We anticipate that our picture provides an alternative view to the physics behind SRS microscopy and facilitate its development and applications.

\appendix

\section{Derivation of effective depth of focus}
We previously estimated the depth of focus by Gaussian fitting of the curvature of the intensity pattern along the optical axis\cite{rigneault2022sensitivity}. Here we derive the depth of focus from the viewpoint of mode density.

The real-space distribution of the complex amplitude of a focused beam can be expressed as the superposition of plain waves as
\begin{equation}
\psi(x,y,z)=\iiint_{S_\mathrm{NA}}\exp[i(k_xx+k_yy+k_zz)]\mathrm d k_x\mathrm dk_y\mathrm dk_z,
\end{equation}
where $S_\mathrm{NA}$ is a part of a spherical surface with a radius of $k_0=2\pi n/\lambda$, which satisfies $\sqrt{k_x^2+k_y^2}\leq k_0\sin\theta$. We can show that 
\begin{equation}
    \psi(0,0,z)\propto\int_{k_{z_0}}^{k_0}\exp(ik_z z)dk_z,
\end{equation}
where $k_{z_0}=k_0\cos\theta \sim k_0(1-(\mathrm{NA}/n)^2/2)$. This means that the focal pattern is the inverse Fourier transform of a rectangular function, i.e. a sinc function. Since $k_z$ is confined from $k_{z_0}$ to $k_0$, the sampling theorem tells us that we can reconstruct the field by sampling the real space along the $z$ axis by a period of 
\begin{equation}
\begin{aligned}
    \frac{2\pi}{k_0-k_{z_0}}&=\frac{2\pi}{k_0(\mathrm NA/n)^2/2}\\
    &=\frac{2n\lambda}{\mathrm{NA}^2}.
\end{aligned}
\end{equation}
This was used as an estimate of $\Delta z$.

\section{Derivation of effective area of focus}

We consider the number of modes in a volume $V=L^3$ of a cube with a width of $L$. We assume a refractive index in the volume is $n$. The sampling theorem tells us that the spatial frequency spectrum in the limited length of $L$ can be perfectly captured by sampling the Fourier space with a spacing of $2\pi/L$, and each sampled point corresponds to a single spatial mode. Thus the number of modes in a limited range of $k$ from $k$ to $k+\mathrm dk$ is given by 
\begin{equation}
    N=\frac{4\pi k^2 \mathrm dk}{(2\pi/L)^3}=\frac {Vk^2}{2\pi^2}\mathrm dk=\frac{2V}{(\lambda/n)^2(c/n)}\mathrm d\omega,
\end{equation}
where we used $k=2\pi n/\lambda$ and $\mathrm dk=n\mathrm d\omega/c$. The sampling theorem tells us that in the frequency interval of $\mathrm d\omega$ can be perfectly captured by sampling with an time interval of $2\pi/\mathrm d\omega$. This means that the number of time-frequency modes in a duration $T$ is $N_\mathrm t = T\mathrm d\omega / 2\pi$. Therefore we obtain
\begin{equation}
    N= \frac{4\pi V}{(\lambda/n)^2(c/n)(T/N_\mathrm t)}.
\end{equation}
If we limit the solid angle of the spatial modes to $\Omega_\mathrm {OB}$, the number of modes becomes
\begin{equation}
    N_\Omega=\frac{\Omega_\mathrm{OB} V}{(\lambda/n)^2(c/n)(T/N_\mathrm t)}.
\end{equation}
This can be used to calculate the number of modes per unit area per unit time given as
\begin{equation}
    \frac{N_\Omega (c/n)}{V}=\frac{\Omega_\mathrm{OB}}{(\lambda/n)^2(T/N_\mathrm t)}.
\end{equation}
This should be equal to $1/S(T/N_\mathrm t)$, where $S$ is the focal area. By using eq. (2), we obtain
\begin{equation}
    S = \frac{(\lambda/n)^2}{\Omega_\mathrm{OB}}\simeq \frac{(\lambda/n)^2}{\pi(\mathrm{NA}/n)^2}=\frac 1 \pi\left(\frac{\mathrm{NA}}{\lambda}\right)^2.
\end{equation} 
This is two times larger than that derived previously\cite{rigneault2022sensitivity}, i.e., $S_\mathrm{prev}=(1/2\pi)(\mathrm {NA}/\lambda)^2$. The difference comes from the fact that in the previous derivation we estimated the focal area by Gaussian fitting of the curvature of the intensity pattern of the focused beam at the focus plane, which does not take into account the intensity spread in outer rings of the focused pattern.

\bibliography{aipsamp}

\end{document}